\documentclass[12pt]{spieman}  
\usepackage{amsmath,amsfonts,amssymb}
\usepackage{graphicx}
\usepackage{setspace}
\usepackage{tocloft}
\usepackage{siunitx}
\usepackage{comment} 
\usepackage{braket}

\title{Group index matched frequency conversion in lithium niobate on insulator waveguides}

\author[a,*]{Pawan Kumar}
\author[a]{Mohammadreza Younesi}
\author[a]{Sina Saravi}
\author[a,b]{Frank Setzpfandt}
\author[a,b]{Thomas Pertsch}
\affil[a]{Institute of Applied Physics, Abbe Center of Photonics, Friedrich Schiller University Jena, Albert-Einstein-Strasse 15, 07745 Jena, Germany}
\affil[b]{Fraunhofer Institute for Applied Optics and Precision Engineering, Albert-Einstein-Str. 7, 07745 Jena, Germany}

\cftpagenumbersoff{figure}
\cftpagenumbersoff{table} 
\begin{document} 
\maketitle

\begin{abstract}
Sources of spectrally engineered photonic states are a key resource in several quantum technologies. Of particular importance are the so-called factorizable biphoton states which possess no spectral entanglement and hence, are ideal for heralded generation of high-purity single photons. An essential prerequisite for generating these states through nonlinear frequency conversion is the control over the group indices of the photonic modes of the source. Here, we show that thin-film lithium niobate on insulator (LNOI) is an excellent platform for this purpose. We design and fabricate periodically poled ridge waveguides in LNOI to demonstrate group index engineering of its guided photonic modes and harness this control to experimentally realize on-chip group index matched type-II sum-frequency generation (SFG) and photon-pair creation through spontaneous parametric down-conversion (SPDC). Also, we numerically study the role of the top cladding layer in tuning the dispersion properties of the ridge waveguide structures and reveal a distinctive difference between the air and silica-clad designs which are currently among the two most common device cladding configurations in LNOI. We expect that these results will be relevant for various classical and quantum applications where dispersion control is crucial in tailoring the nonlinear response of the LNOI-based devices.                        
\end{abstract}


{\noindent \footnotesize\textbf{*}Pawan Kumar,  \linkable{pawan.kumar@uni-jena.de} }

\begin{spacing}{2}   

\section*{Introduction}
The field of integrated linear and nonlinear optics has seen rapid technological progress based on recent developments in the thin-film  lithium niobate on insulator (LNOI)  photonic platform \cite{Zhang2017, Wang2018_EOM, Saravi2021_AOM}. Improvements in microfabrication techniques have enabled reliable monolithic fabrication of nanoscale low-loss optical waveguide structures with typical cross-sectional dimensions below 1 $\mu m^2$. This promises to truly meet the requirements in high on-chip integration density, high performance and scalability of fabrication \cite{Luke2020}. More crucially, lithium niobate (LN), being a $\chi^{(2)}$-nonlinear material, allows an array of integrated nonlinear optical applications such as classical frequency conversion \cite{Geiss2015, Wang2018}, on-chip frequency comb creation \cite{Zhang2019_EO_comb}, supercontinuum generation, \cite{Yu2019, ReigEscal2020} and photon-pair generation with strong spectral entanglement and high generation rates \cite{Zhao2020_PRL} through efficient spontaneous parametric down-conversion (SPDC). 

Apart from high conversion efficiencies\cite{Wang2018, Rao2019}, an often needed capability in classical and quantum nonlinear frequency conversion processes is control over the spectral characteristics of the generated light. This is especially relevant for quantum applications requiring generation of spectrally tailored quantum states. A preeminent example of such an application is the heralded generation of high purity single-photon states through SPDC. In SPDC, a higher energy pump (P) photon at frequency $\omega_\mathrm{P}$ spontaneously splits into a pair of lower energy photons, called signal (S) and idler (I) with frequencies $\omega_\mathrm{S}$ and $\omega_\mathrm{I}$ inside a $\chi^{(2)}$-medium, such that  $\omega_\mathrm{P} = \omega_\mathrm{S} + \omega_\mathrm{I}$. If the SPDC process is engineered to eliminate the spectral entanglement between the two photons, such that the generated two-photon joint state is spectrally factorizable, then the generation of a spectrally pure single-photon state can be heralded by detecting the other photon of the pair\cite{uRen2005generation, mosley2008heralded}.  
 
Spectral engineering of SPDC to produce such factorizable states requires careful manipulation of the group indices (n$_g$) of pump, signal and the idler modes such that either the condition $ \mathrm{n}^\mathrm{S}_g \leq \mathrm{n}^\mathrm{P}_g \leq \mathrm{n}^\mathrm{I}_g $ or $ \mathrm{n}^\mathrm{I}_g \leq \mathrm{n}^\mathrm{P}_g \leq \mathrm{n}^\mathrm{S}_g $ can be fulfilled\cite{Grice2001}. In general achieving such control over the dispersion properties of a $\chi^{(2)}$ nonlinear source at wavelengths of interest is not always possible. 

Ridge waveguides in LNOI allow broad control over the dispersion of its optical modes\cite{Jankowski2020}. This is in contrast to the conventional low-index contrast in-diffused waveguides in LN where the dispersion properties of the modes are mostly dominated by that of the bulk LN material. In the case of LNOI, tight modal confinement of light in subwavelength dimension waveguiding structures can lead to dispersion engineering opportunities. Through geometry-induced dispersion control, LNOI ridge waveguides can allow manipulation of group velocity and group-velocity dispersion of the modes. In recent works, this dispersion engineering capability has received special attention \cite{Jankowski2021disper}. Through appropriate waveguide design, dispersion parameters have been optimized to significantly enlarge the bandwidth of second harmonic generation and SPDC processes compared to that in the bulk medium \cite{Jankowski2020, ledezma2021intense, Javid2021, solntsev2011cascaded}.  

So far most of the experimental efforts towards dispersion engineering in LNOI nanophotonics have been focused on the so-called type-0 processes to exploit its highest nonlinear tensor element, d$_{33}$. 
However, a type-0 process restricts the range of possibilities in dispersion control since it does not fully leverage the strong birefringence of ridge waveguides. 
In particular, to satisfy the factorizability condition for generation of frequency-degenerate signal and idler photons, the two photons have to be generated in two different modes to first realize different group indices for them, and then try to engineer the pump group index to a value between them.
For this scenario, a type-II process is needed, which utilizes the $d_{31}$ element of the nonlinear tensor and involves two orthogonally polarized modes for the signal and idler photons: the transverse magnetic (TM) and transverse electric (TE) modes. In fact, it has been theoretically shown that by exploiting the geometry-dependent birefringence of the modes in LNOI ridge waveguides, the group indices can be tailored to satisfy the condition for generation of factorizable photon-pairs \cite{Kang2014}. 

In this work, we experimentally demonstrate the dispersion-engineering capability of LNOI ridge waveguides for spectral control of type-II frequency conversion processes. Specifically, we show that in a type-II process, the group indices of the signal and pump modes can be made equal, $\mathrm{n}^\mathrm{P}_g = \mathrm{n}^\mathrm{S}_g $, through a suitable choice of the design parameters.
We verify this experimentally by mapping the spectral properties of the type-II process in sum-frequency generation (SFG) experiments, where the waveguide is excited at the signal and idler wavelengths and pump photons are generated through the SFG process. 
This shows the capability of LNOI waveguides for dispersion engineering of type-II frequency conversion. The  group index matching demonstrated here is highly relevant for realizing on-chip sources of spectrally pure single photons as it is required for the  generation of factorizable photon-pair states through SPDC\cite{mosley2008heralded}. 
Since modal-confinement is crucial in controlling the group indices of the modes, we also perform a numerical analysis of the effect of top-cladding layer and waveguide dimensions on group index. It is shown that while air-clad higher-index contrast ridge waveguides offer more diverse control, the silica-clad waveguides can offer higher robustness to device dimension variations. Finally, we generate photon pairs through continuous-wave SPDC to highlight the potential of our tested systems for quantum technologies.      

\section*{Phase-matching function}
To study the effect of dispersion engineering on the spectral features of type-II frequency conversion, we start by describing the corresponding SPDC process. The biphoton state $ \ket{\psi} $ generated in SPDC is described in the frequency domain as $ \ket{\psi} = \iint d\omega_\mathrm{S} d\omega_\mathrm{I} J(\omega_\mathrm{S}, \omega_\mathrm{I}) \ket{1}_\mathrm{S} \ket{1}_\mathrm{I}$, where the joint spectral amplitude (JSA), $J(\omega_\mathrm{S}, \omega_\mathrm{I})$, contains information about spectral entanglement between the signal-idler pair produced from the pump photon\cite{Grice2001}. The JSA can be shown to be $ J(\omega_\mathrm{S}, \omega_\mathrm{I}) \propto  A_p(\omega_\mathrm{P} = \omega_\mathrm{S} + \omega_\mathrm{I})\times \phi (\omega_\mathrm{S} , \omega_\mathrm{I})$ with $ A_p(\omega_\mathrm{P}) $ being the pump photon spectral amplitude and $ \phi (\omega_\mathrm{S} , \omega_\mathrm{I}) $ being the phase matching function (PMF). If the factorizability condition on the group indices of the photonic modes is fulfilled, the shape of the resulting PMF enables the generation of a spectrally factorizable state with  $ J(\omega_\mathrm{S}, \omega_\mathrm{I}) =  u(\omega_\mathrm{S})\times v(\omega_\mathrm{I})$. 

The PMF for a quasi-phase-matched (QPM) type-II frequency conversion process depends on the phase-mismatch vector $\Delta K$ and length $L$ of the waveguide, and is given by $ \phi = \mathrm{sinc}(\Delta K L/2) \times$ $ \mathrm{exp}(-\mathrm{i} \Delta K L/2)$, where $\Delta K (\omega_\mathrm{S}, \omega_\mathrm{I}) = K^\mathrm{S}(\omega_\mathrm{S}) + K^\mathrm{I}(\omega_\mathrm{I}) - K^\mathrm{P}(\omega_\mathrm{P}) + (2\pi/\Lambda)$. Here, $K^{i= \mathrm{P}, \mathrm{S}, \mathrm{I}}$ are the wave vector of the respective modes of the waveguide. For QPM assisted frequency conversion the nonlinearity of the waveguide is periodically switched in direction with a period of $\Lambda$ along its length though a technique called periodic poling. The resulting nonlinear grating vector $ (2\pi/\Lambda)$ compensates for the mismatch of the effective indices of the waveguide modes leading to efficient frequency conversion. QPM, which is a well established technique in bulk LN, has recently been adapted to the LNOI thin-film platform \cite{Wang2018, Rao2019,Younesi2021}. This decouples the phase-matching requirement from the dispersion control such that the geometry of the waveguide can be used to control the group velocities of the modes to modify the spectral dependence of the PMF while the central phase-matching wavelengths can be fixed through the choice of the QPM period $\Lambda$ so that $ \Delta K = 0$.         

\begin{figure}
\centering
\includegraphics[width=0.8\linewidth]{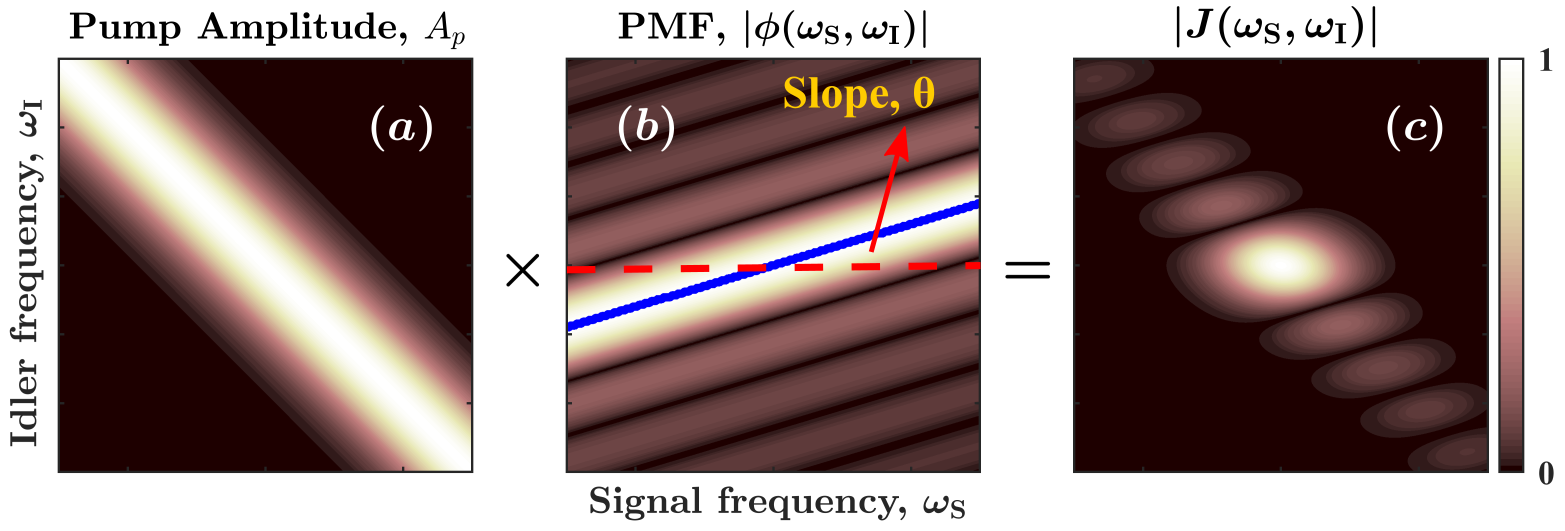}
\caption{Schematic of the JSA, $ J = A_p \times \phi $, of a biphoton state. (a) Gaussian pump spectral amplitude, $A_p$ (b) PMF, $\phi$, of a type-II frequency conversion process. The blue solid line, referred to as the PMC, is oriented at an angle of $\theta$ with the $\omega_\mathrm{S}$-axis (represented schematically by the horizontal red dashed line) and represents the maximum of PMF. (c) Resulting JSA of the signal-idler pair.} 
\label{fig:No0}
\end{figure}

Since the quantum process of SPDC and its reverse classical process of SFG share the same PMF, by virtue of the quantum-classical correspondence principle\cite{Lenzini2018} , the experimental investigation of the PMF for SPDC is also possible by studying the corresponding SFG process. This immensely reduces the experimental effort in probing the spectral dependence of the PMF. The SFG intensity, $I_\mathrm{SFG}$, is related to $\phi (\omega_\mathrm{S} , \omega_\mathrm{I})$ as $ I_\mathrm{SFG} \propto |\phi (\omega_\mathrm{S} , \omega_\mathrm{I})|^2  = \mathrm{sinc}^2(\Delta K L/2)$. Vanishing $\Delta K  $ maximizes the sinc function and leads to the largest SFG intensity. Thus $\Delta K(\omega_\mathrm{S}, \omega_\mathrm{I}) = 0 $ gives the relation between the signal and idler frequency defining the phase-matching-curve (PMC) 
of the SFG and SPDC processes. The slope of the PMC is related to the group indices as\cite{uRen2005generation}
\begin{equation} \label{slope}
    \theta = \mathrm{tan}^{-1}\left\{ -\left( \mathrm{n}^\mathrm{P}_g  - \mathrm{n}^\mathrm{S}_g\right)/ \left( \mathrm{n}^\mathrm{P}_g  - \mathrm{n}^\mathrm{I}_g\right) \right \},
\end{equation}
thus directly revealing the effect of group-index engineering on the spectrum of biphoton state generated in SPDC. We depict this schematically in Fig.~\ref{fig:No0}. The pump amplitude $ A_p(\omega_\mathrm{p}) $, shown in Fig.~\ref{fig:No0}(a), is oriented at a fixed angle of $-45^{\circ}$ with respect to the $ \omega_\mathrm{S}$-axis due to energy conservation, $\omega_\mathrm{p} = \omega_\mathrm{S} + \omega_\mathrm{I}$. On the other hand, the slope of the PMC can be tuned through dispersion control to lie in the range $ 0^{\circ} \leq \theta \leq 90^{\circ} $ as is required to reach the  factorizability condition.   This is shown in Fig.~\ref{fig:No0}(b) by the blue solid line overlaid on the two-dimensional plot for the PMF $\phi (\omega_\mathrm{S} , \omega_\mathrm{I})$. The resulting JSA  $J(\omega_\mathrm{S}, \omega_\mathrm{I})$ of the signal-idler pair is displayed in Fig.~\ref{fig:No0}(c).  It is composed of a central circular-looking high intensity lobe surrounded by several smaller intensity lobes along the $-45^{\circ}$ diagonal.                   

The factorizability of  $J(\omega_\mathrm{S}, \omega_\mathrm{I})$ can be optimized for a given PMF by choosing an appropriate spectral bandwidth of the pump\cite{Grice2001}. In general, the extent of this optimal factorizability and consequently, the purity of the heralded single photon states is larger for the two extreme cases when the slope of PMC is either $\theta = 0^{\circ}$ or $ \theta = 90^{\circ}$ compared to other slopes in the range $ 0^{\circ} \leq \theta \leq 90^{\circ}$\cite{Jankowski2021disper}. These specific cases can be achieved through group index matching, $ \mathrm{n}^\mathrm{P}_g = \mathrm{n}^\mathrm{S}_g \neq \mathrm{n}^\mathrm{I}_g$ and $ \mathrm{n}^\mathrm{P}_g = \mathrm{n}^\mathrm{I}_g \neq \mathrm{n}^\mathrm{S}_g $ respectively.  Here, we demonstrate the first case, where the group indices of the pump and signal are nearly matched while that of the idler is smaller, $ \mathrm{n}^\mathrm{P}_g \approx \mathrm{n}^\mathrm{S}_g > \mathrm{n}^\mathrm{I}_g $, leading to a PMC slope of $\theta \approx 0^{\circ}$.                                 

\section*{Group-index matching} 
\begin{figure}
\centering
\includegraphics[width=0.5\linewidth]{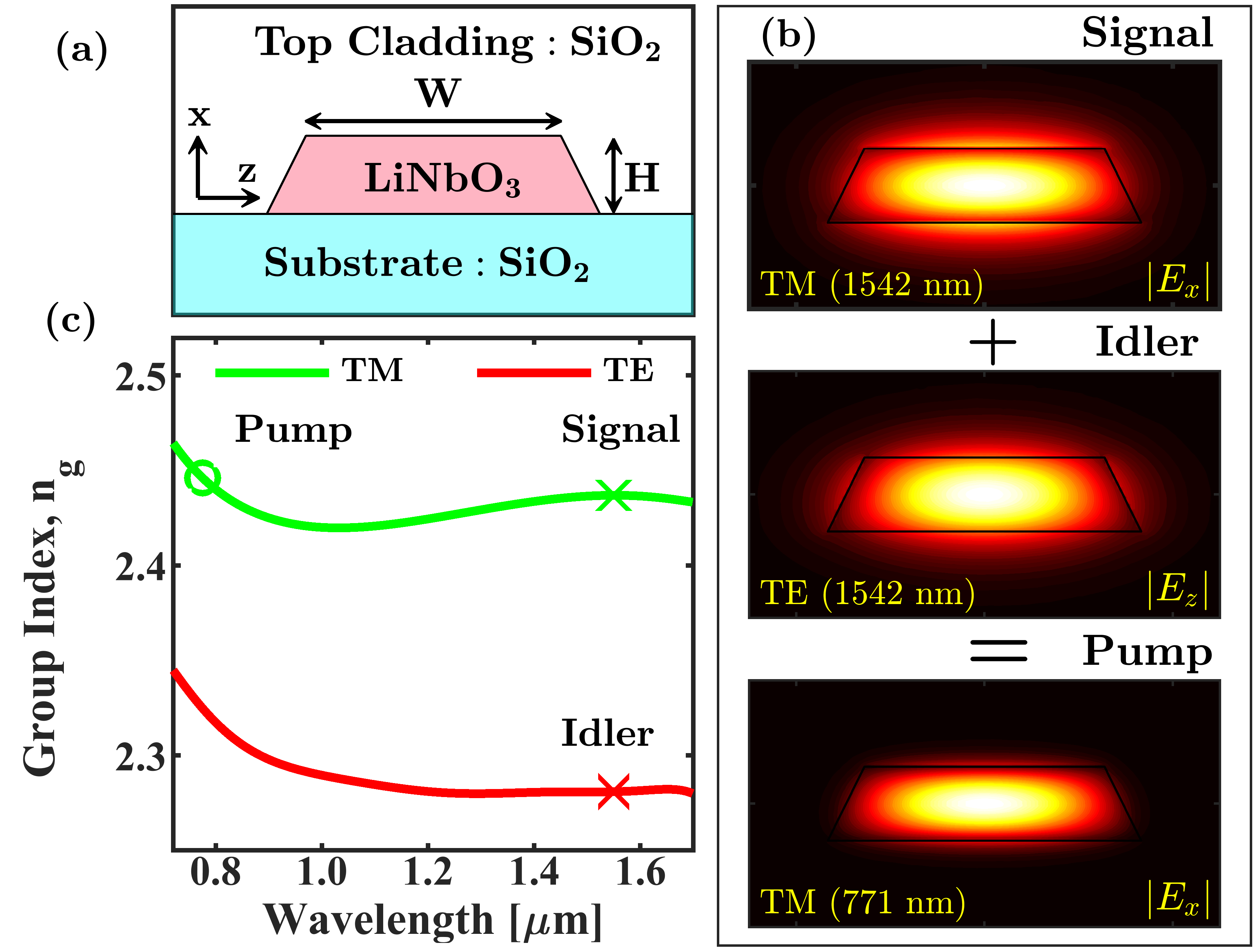}
\caption{(a)  Schematic of the LNOI ridge waveguide in x-cut thin-film with top width $ \mathrm{W} = 1275 $ nm, height $ \mathrm{H} = 600 $ nm and base side-wall angle of 72$^{\circ}$. (b) The three fundamental modes of the waveguide involved in type-II SFG. (c) Variation of group index with wavelength for TE and TM modes.  Group indices of pump and signal are nearly matched.} 
\label{fig:No1}
\end{figure}

We consider an x-cut LNOI thin-film geometry as shown in Fig.~\ref{fig:No1}(a) for designing the ridge waveguide. The waveguide has a height of $\mathrm{H}= 600$ nm and is cladded from the top with a thick silica layer. The top with of this waveguide is $\mathrm{W}= 1275$ nm. Its three fundamental modes involved in type-II frequency mixing are shown in Fig.~\ref{fig:No1}(b). To phase match the process we used a QPM period of $\Lambda = 3 \ \mu m$ which results in phase matching of the signal and idler waves at the central wavelength of  $\lambda_\mathrm{S} = \lambda_\mathrm{I} = 1542 $ nm to generate the pump at half of their wavelength. The length of the waveguide is $L = 4$ mm.            

Fig.~\ref{fig:No1}(c) shows the dependence of the group index $\mathrm{n}_g$ of TE and TM modes of the waveguide on the wavelength. The green curve belongs to the TM mode on which we have marked the position of signal and pump by green cross and circle respectively. The corresponding group index $\mathrm{n}^\mathrm{I}_g$ of the idler is highlighted on the red curve belonging to the TE mode by a red cross. We note that the green curve shows an interesting behaviour, such that it first decreases with wavelength till about $\lambda \approx 1 \mu m$ and then shows an increase. As a result $\mathrm{n}^\mathrm{S}_g$ and $\mathrm{n}^\mathrm{P}_g$ can be matched. For the design presented here, we calculate the group indices of pump and signal to be $\mathrm{n}^\mathrm{P}_g = 2.447$ and $\mathrm{n}^\mathrm{S}_g = 2.437$, while $\mathrm{n}^\mathrm{I}_g = 2.281$ is significantly smaller. Using these values we estimate the slope of the PMC using Eq.~(\ref{slope}) to be $ \theta = - 3.45^{\circ}$.    

\begin{figure}
\centering
\includegraphics[width=0.7\linewidth]{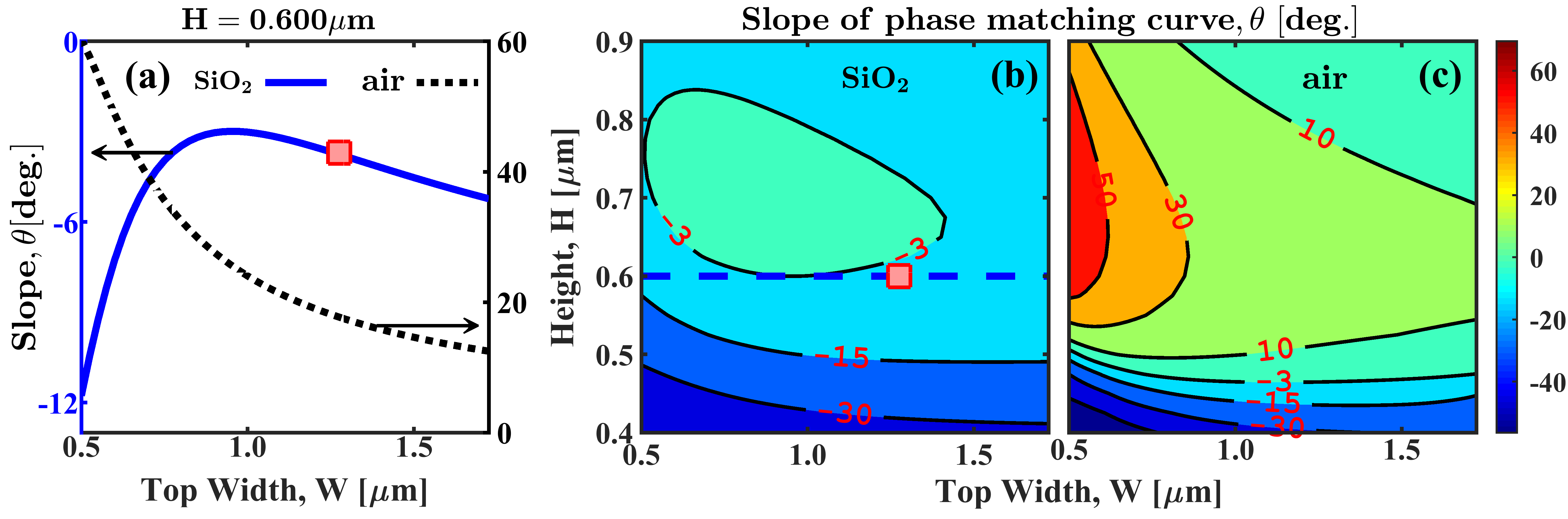}
\caption{Slope $\theta$ of PMC for type-II degenerate-frequency SFG with the fundamental wavelength of 1542 nm. (a) Variation of the slope with waveguide width $\mathrm{W}$ for a fixed height $\mathrm{H}$ for the silica cladding (blue solid curve) and air cladding (black dotted curve) cases. The red square denotes the waveguide of width $\mathrm{W} = 1275$ nm.  (b)-(c) Dependence of PMC slope on height and width of waveguide for silica and air cladding designs respectively. The blue dashed line in (b) indicates the silica-clad waveguides of constant height $\mathrm{H} = 600$ nm.            
} 
\label{fig:No2}
\end{figure}

We now look at the influence of the waveguide dimensions on the group indices with an aim to investigate the tunability of the PMC slope. We consider waveguides with silica and air top cladding since the index-contrast between the core and surroundings is significantly different in these cases, which affects their dispersion properties. In Fig.~\ref{fig:No2}(a) we show the variation of the slope $ \theta $ with the top width of the waveguide for a fixed height. The central signal and idler phase-matching wavelength is assumed to be equal to 1542 nm in each case for calculating the slope. The blue solid curve represents the silica-clad case where the minimal slope of $ \theta \approx - 3^{\circ}$ is attained for a waveguide width of $ \mathrm{W}= 950$ nm. The red square on this curve represent the design we discussed before with a width of  $ \mathrm{W}= 1275$ nm. We note that although this is not the optimal case, the difference between their slopes is small. On the other hand, the air-clad case displayed by the black dotted curve shows higher sensitivity to the waveguide width and thus allows wider tunability of the slope. Remarkably, $ \theta $ in the air-clad case is positive and has a larger magnitude meaning that $ \mathrm{n}^\mathrm{S}_g $ is substantially greater than $\mathrm{n}^\mathrm{P}_g $ in contrast to the silica cladding case. The contour plots in Fig.~\ref{fig:No2}(b) and (c) show the dependence of the slope on the height and width of the waveguide for silica and air-clad cases respectively. Comparing these shows, that the silica-clad design can offer a more robust way for group index matching with $\theta \approx 0$, since there exists a globally optimal waveguide design with dimensions $ \mathrm{W}= 775$ nm and $ \mathrm{H}= 700$ nm. In this case, sensitivity to variations in the device dimensions or the wavelength of operation is quite low compared to the air-clad case.

\section*{SFG experiments} 
We experimentally confirm group-index matching by performing SFG experiments. This is done for four waveguides having widths $ \mathrm{W}$ in the range 1275 nm to 1380 nm and the same height $ \mathrm{H}= 600$ nm. They have silica cladding on the top. Since their QPM period was the same, $\Lambda = 3 \ \mu m$, their central signal-idler phase-matching wavelengths are different.   

For SFG experiments we combine two independent continuous-wave  tunable lasers using a fiber-based polarization beam combiner and  couple them simultaneously into the waveguide with the help of a lensed tip single mode fiber. Their polarization state is controlled individually on the input side using two fiber-based polarization controllers so that one of them excites the fundamental TM mode (signal) while the other excites the TE mode (idler) of the waveguide. The waveguide chip and the lensed fiber are placed on 3-axis translation stages to optimize the coupling. On the output side, a broadband collecting objective was placed to collect the signal and idler beams and the sum-frequency radiation (pump). The lower wavelength SFG beam is then separated from the higher wavelength fundamental beams using a longpass dichroic beamsplitter whose pass edge is around 1 $\mu m$. Finally, the signal and idler beams, which are in orthogonal polarization, are separated using a polarization beamsplitter and the power of three beams is measured using appropriate power meters. 

\begin{figure}
\centering
\includegraphics[width=0.6\linewidth]{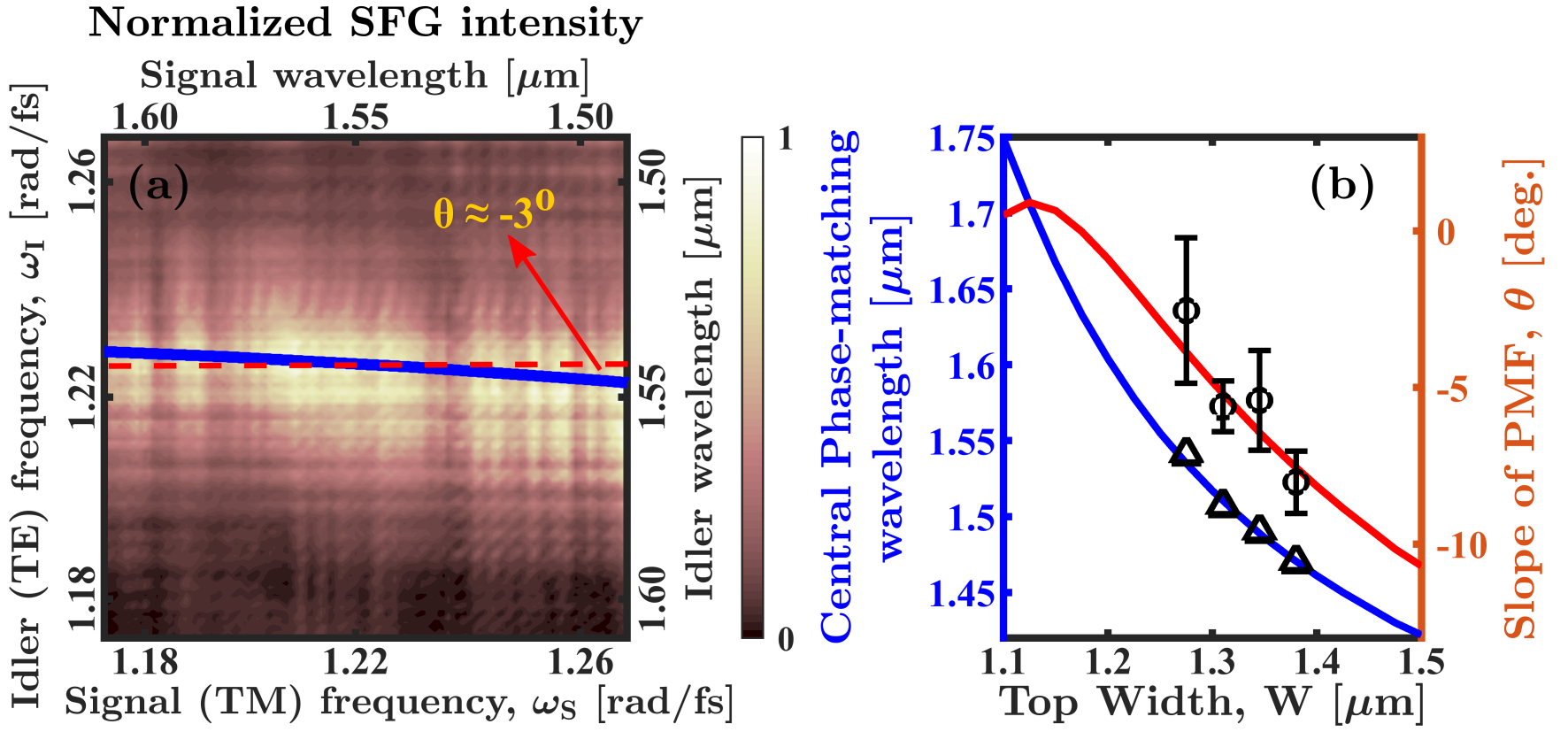}
\caption{ (a) Normalized SFG intensity from the waveguide of width $\mathrm{W}= 1275$ nm for varying signal and idler wavelengths. The solid blue curve is the theoretical PMC and $\theta$ denotes its slope. (b) Central phase-matching wavelength (blue curve) and the corresponding slope of PMC (red curve) for waveguides of different widths. A constant QPM period of $\Lambda = 3 \ \mu m$ is assumed for all waveguides.} 
\label{fig:No3}
\end{figure}

We present the result of such a measurement for the waveguide of width $ \mathrm{W} = 1275 $ nm as a two-dimensional plot in Fig.~\ref{fig:No3}(a). This is obtained by sweeping the wavelength of the lasers from 1490 nm to 1610 nm. The normalized SFG intensity essentially represents the PMF $ \phi(\lambda_\mathrm{S},  \lambda_\mathrm{I})$ such that its maximum maps out the PMC of the SFG process. We can see that the SFG intensity produces a band which is oriented nearly flat along the horizontal direction parallel to the signal frequency axis. This clearly shows that the signal and pump are group index matched. The solid blue curve overlaid on the SFG plot is the numerically calculated PMC while the dashed red line is a horizontal line parallel to the $ \omega_\mathrm{S}$ axis.  The slope of the PMC is quite small, being equal to $\theta \approx -3^{\circ} $.

To obtain the value of the slope $\theta$ from experiments, we first determine the experimental PMC. This is extracted from the SFG experimental data by determining the idler wavelength which maximizes the SFG intensity for each signal wavelength. Then a linear function is fitted to the experimental PMC to get its slope $\theta$. In Fig.~\ref{fig:No3} (b) we show the result of such analysis for the four waveguides presented in this work. The solid blue curve in the figure shows the central signal-idler phase-matching wavelength for SFG for waveguides of different widths. The black triangles marked on it correspond to the different waveguides studied experimentally. The theoretical slope of their PMC is shown in the figure by the solid red curve. The black circles on this curve denote the experimentally observed slope $\theta$ obtained from SFG measurements. We see that with a increase in waveguide width the slope of the PMC grows in magnitude. However, its variation is small such that increasing the width from $ \mathrm{W} = 1100$ nm to 1500 nm changes $\theta$ from about $ 1^{\circ} $ to  $ -10^{\circ} $. Remarkably, the central phase-matching wavelength changes substantially with increasing width. Such dispersion characteristics allow the central wavelength of nonlinear frequency conversion to be easily tuned while still approximately maintaining the group-index matching through appropriate waveguide design.                                         

We point out that the experimental SFG intensity  shown in Fig.~\ref{fig:No3}(a) differs from the well-known $\mathrm{sinc}^2$ shape of the PMF. This is due to the presence of loss. We  ascertain the magnitude of these losses for signal and idler by performing Fabry-Perot interference measurement \cite{Hu2009}. From the visibility of fringes in transmission of spectral intensity, we estimate these to be 20 dB/cm and 26 dB/cm, for idler and signal respectively. Lossy guiding modes of the waveguide also affect the conversion efficiency of SFG. The normalized conversion efficiency, which is defined as $\eta = P_\mathrm{P}/\left( P_\mathrm{S} P_\mathrm{I} L^2\right)$, is estimated to be about 560 \% $W^{-1} \text{cm}^{-2}$ if the waveguides were lossless. Here, $P_{\mathrm{P}, \mathrm{S}, \mathrm{I}}$ are the respective powers of the three modes. The experimentally observed efficiency, however, is only 4 \% $W^{-1} \text{cm}^{-2}$. The theoretically estimated $\eta$ assuming a lossless waveguide agrees with the observed efficiency when we take into account the optical losses of the modes, their individual collection efficiencies and limited transmission through the collecting objective. 

\section*{Continous-wave SPDC} 
To highlight the usefulness of dispersion engineered LNOI waveguides for quantum applications, we performed a simple SPDC experiment in the waveguide of width $\mathrm{W}= 1275$ nm. For this, a continuous-wave laser operating at the pump wavelength of $\lambda_ \mathrm{P}$ = 770 nm and input power of 20 \text{mW} was coupled into the TM mode of the waveguide to drive the type-II SPDC process. The generated signal-idler photon pairs were collected using a broadband objective lens and the pump was filtered out using several longpass filters. We then couple the signal and idler photons into single-mode fibers and send them to two superconducting nanowire single-photon detectors (SNSPD - Single Quantum EOS). We delay the idler photon with respect to the signal and record the coincidences between the two SNSPD detection events using a  time-to-digital converter (IDQuantique ID800). To ascertain whether two events are coincident or not, a temporal coincidence window of 162 ps duration is chosen. 

\begin{figure}
\centering
\includegraphics[width=0.30\linewidth]{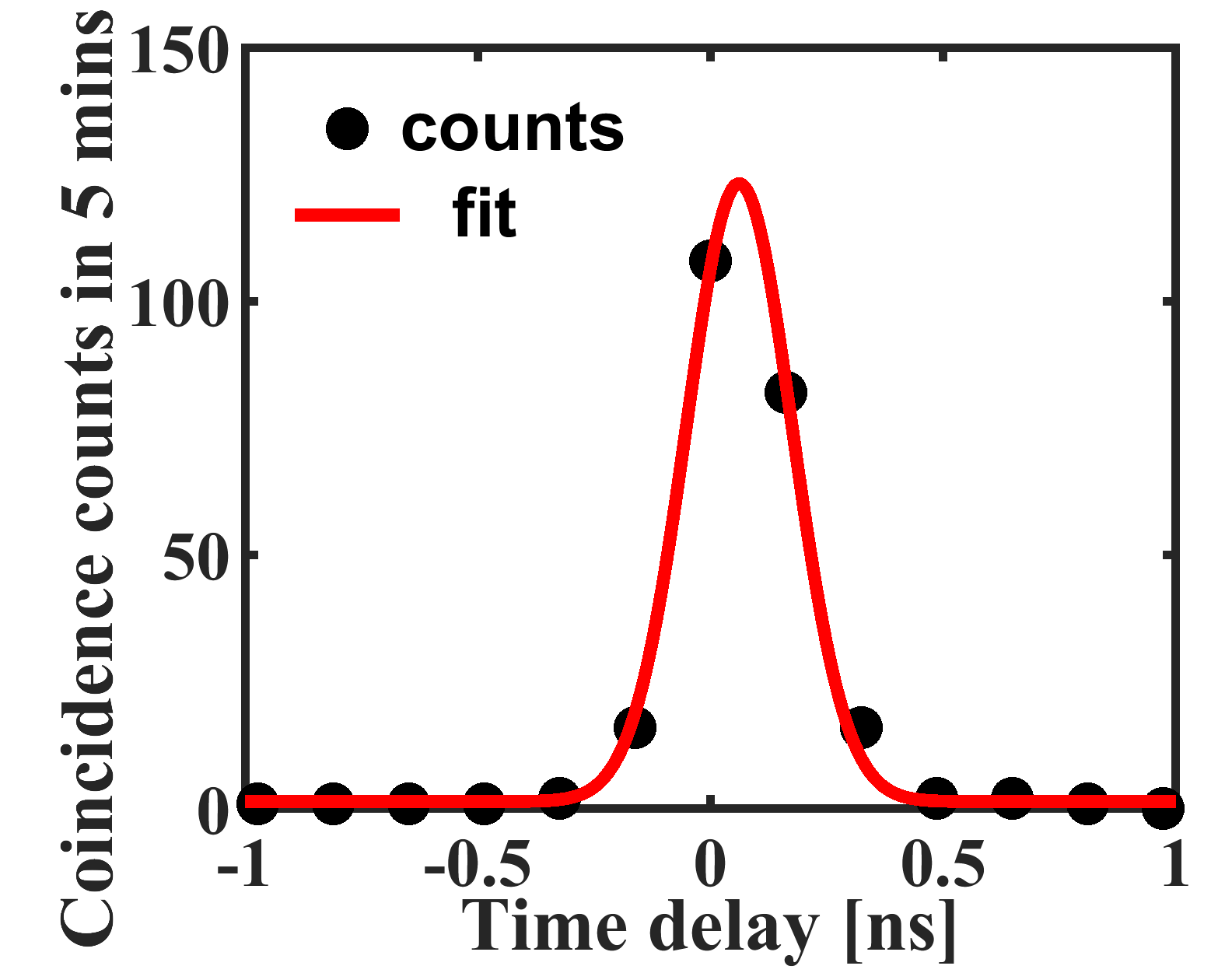}
\caption{Coincidence measurement on signal and idler photons generated by type-II SPDC.}
\label{fig:No4}
\end{figure}

The observed coincidence counts as a function of time delay between idler and signal are shown in Fig.~\ref{fig:No4}. A coincidence peak with a coincidence to accidental ratio (CAR) of about 85 is observed. This confirms the generation of temporally correlated photon pairs in the waveguide through SPDC. We also confirmed that idler and signal photons are in orthogonal polarization as expected for a type-II SPDC process. 
We expect the magnitude of CAR to improve in future versions of the device fabricated through a more optimised fabrication process.  

\section*{Conclusions}
In conclusion, we have presented the concept of group-index matching for thin-film LNOI ridge waveguides and experimentally demonstrated its application in tuning the phase-matching curve of a type-II SFG process. This is highly relevant for generation of factorizable biphoton states to realize heralded sources of spectrally pure single photons. We show that waveguide dimensions together with the choice of the top cladding layer determine the extent of control over group indices of waveguide modes. Finally, we demonstrate photon-pair generation through SPDC in the waveguide to highlight its application in quantum technologies.

\subsection*{Disclosures}
The authors declare no conflict of interest. 
\subsection* {Acknowledgments}
The authors acknowledge funding from the German Federal Ministry of Education and Research (BMBF) under the project identifiers 13N14877 (QuanIm4Life), 13N16108 (PhoQuant); from the German Research Foundation (DFG) under the project identifiers PE 1524/13-1 (NanoPair), SE 2749/1-1 (NanoSPDC), 398816777-SFB 1375 (NOA); and from the Thuringian Ministry for Economic Affairs, Science and Digital Society under the project identifier 2021 FGI 0043 (Quantum Hub Thuringia).

\bibliography{article}   
\bibliographystyle{spiejour}   

\listoffigures
\end{spacing}
\end{document}